\documentclass[11pt]{article}

\usepackage{acl}

\usepackage{times}
\usepackage{latexsym}

\usepackage[T1]{fontenc}
\usepackage[utf8]{inputenc}

\usepackage{microtype}

\usepackage{inconsolata}

\usepackage{graphicx}

\usepackage{amsmath}
\usepackage{amsfonts}
\usepackage{multirow}
\usepackage{algorithm}
\usepackage{algpseudocode}

\title{Anchored Cyclic Generation: A Novel Paradigm for Long-Sequence Symbolic Music Generation}

\author{
  \textbf{Boyu Cao}\textsuperscript{1\thanks{Equal contribution.}} \quad
  \textbf{Lekai Qian}\textsuperscript{1\footnotemark[1]} \quad
  \textbf{Dehan Li}\textsuperscript{1} \quad
  \textbf{Haoyu Gu}\textsuperscript{1} \quad
  \textbf{Mingda Xu}\textsuperscript{1} \quad
  \textbf{Qi Liu}\textsuperscript{1\thanks{Corresponding author.}} \\[6pt]
  \textsuperscript{1}South China University of Technology, School of Future Technology \\
  \texttt{drliuqi@scut.edu.cn}
}

\begin{document}
\maketitle

\begin{abstract}
Generating long sequences with structural coherence remains a fundamental challenge for autoregressive models across sequential generation tasks. In symbolic music generation, this challenge is particularly pronounced, as existing methods are constrained by the inherent severe error accumulation problem of autoregressive models, leading to poor performance in music quality and structural integrity. In this paper, we propose the Anchored Cyclic Generation (ACG) paradigm, which relies on anchor features from already identified music to guide subsequent generation during the autoregressive process, effectively mitigating error accumulation in autoregressive methods. Based on the ACG paradigm, we further propose the Hierarchical Anchored Cyclic Generation (Hi-ACG) framework, which employs a systematic global-to-local generation strategy and is highly compatible with our specifically designed piano token, an efficient musical representation. The experimental results demonstrate that compared to traditional autoregressive models, the ACG paradigm achieves reduces cosine distance by an average of 34.7\% between predicted feature vectors and ground-truth semantic vectors. In long-sequence symbolic music generation tasks, the Hi-ACG framework significantly outperforms existing mainstream methods in both subjective and objective evaluations. Furthermore, the framework exhibits excellent task generalization capabilities, achieving superior performance in related tasks such as music completion.
\end{abstract}

\section{Introduction}

Autoregressive sequence modeling has achieved remarkable success across various domains, from natural language processing to structured content generation. However, maintaining long-term coherence and structural integrity in extended sequences remains a persistent challenge due to error accumulation during iterative generation. Symbolic music generation exemplifies this challenge: as a core branch of music generation, it produces discrete musical representations with structured, interpretable characteristics \citep{briot2017deep}, yet modeling long-sequence symbolic music has become a primary challenge with the rapid development of deep learning technologies.

Long-sequence symbolic music generation requires maintaining both local coherence and global structural integrity. Autoregressive models, the mainstream approach, predict subsequent segments based on historical content but face significant limitations. Early RNN/LSTM approaches exhibit degraded quality and stylistic drift in longer sequences, making it difficult to maintain long-term consistency. Attention-based methods like Transformers, combined with MIDI event encoding \citep{oore2018time} or ABC notation, perform well on short sequences but face exponentially growing computational complexity \citep{child2019generating} and severe error accumulation as length increases. Diffusion models \citep{mittal2021symbolic} offer an alternative but struggle to generate complete long-sequence music efficiently.

Addressing these challenges, we propose the Anchored Cyclic Generation (ACG) paradigm, which introduces determined musical content as anchors in each generation cycle to recalibrate the generation process, effectively mitigating error accumulation and achieving smooth transitions between musical segments. ACG ensures local musical quality while maintaining structural integrity of long-sequence music, with significant advantages in time complexity. Based on ACG, we further propose a Hierarchical Anchored Cyclic Generation (Hi-ACG) framework with a novel piano token representation. The framework adopts a hierarchical strategy: a sketch loop captures high-level semantic features such as modality, harmonic progression, and overall structure, providing global guidance for subsequent processes; a refinement loop then generates detailed note-level content to ensure local coherence and expressiveness. Experiments demonstrate that Hi-ACG maintains long-term structural and stylistic consistency while achieving precise duration control.

Overall, our contributions are as follows:

\begin{itemize}
    \item We present the ACG paradigm, which significantly mitigates error accumulation in long-sequence generation tasks such as symbolic music modeling. Our method demonstrates improved time complexity and lower computational costs compared to conventional autoregressive approaches.
    \item We present Hi-ACG, a hierarchical framework that generates music from global to local levels. It solves structural integrity problems in long sequences, provides precise duration control, and offers high interpretability.
    \item We propose a Piano Token musical representation---an efficient tokenization method that converts piano roll data into musical tokens. This approach is highly compatible with our Hi-ACG framework while remaining adaptable to other autoregressive symbolic music generation models.
    \item Based on mathematical analysis and experimental validation, our proposed ACG paradigm and Hi-ACG framework effectively mitigate error accumulation, enhance the generation quality and structural integrity of long-sequence symbolic music, and demonstrate superior performance over existing models in both objective and subjective evaluations.
\end{itemize}

\section{Related Work}

\subsection{Symbolic Music Generation}

Symbolic music generation aims to automatically generate discrete musical representations with musicality and interpretability. Early approaches based on rules \citep{ebcioglu1988expert} and statistical modeling \citep{conklin1995multiple} laid the foundation for data-driven methods. With deep learning, RNNs \citep{eck2002finding,sturm2016music}, VAE-based models like MusicVAE \citep{roberts2018hierarchical}, and adversarial approaches like MuseGAN \citep{dong2017museganmultitracksequentialgenerative} became mainstream, though they struggled with long sequences.

Recently, Transformers \citep{vaswani2017attention} and diffusion models \citep{ho2020denoisingdiffusionprobabilisticmodels} have advanced the field. RIPO Transformer \citep{Guo_2023} enhanced melody modeling with novel attention mechanisms. TunesFormer \citep{wu2023tunesformerformingirishtunes} enabled bar-level controlled generation. ChatMusician \citep{yuan2024chatmusicianunderstandinggeneratingmusic} demonstrated how language models can understand and generate music through unified text-music pretraining.

\subsection{Long-Sequence Symbolic Music Modeling}

Long-sequence symbolic music modeling is a key challenge in music generation. Traditional autoregressive methods have several problems like error accumulation, high computational complexity, and vanishing gradients.

The challenge of modeling long sequences is shared across domains. In natural language processing, methods such as Longformer \citep{beltagy2020longformerlongdocumenttransformer} and hierarchical text generation approaches have addressed similar issues of computational complexity and long-range coherence. Our work draws inspiration from these cross-domain advances while addressing the unique structural requirements of symbolic music.

\subsubsection{Transformer-based Methods}

Transformers are widely used in long-sequence symbolic music generation due to their ability to model long-range dependencies. Music Transformer \citep{huang2018musictransformer} first applied Transformer architecture to symbolic music generation using relative positional encoding. Longformer \citep{beltagy2020longformerlongdocumenttransformer} introduced sparse attention to reduce computational cost. Museformer \citep{yu2022museformertransformerfinecoarsegrained} proposed structure-aware FC-Attention using bar-level summary tokens and multi-scale attention. Compound Word Transformer \citep{hsiao2021compoundwordtransformerlearning} introduced compound token representation. BPE-Music \citep{fradet2023bytepairencodingsymbolic} employed subword modeling to compress token sequences. MuPT \citep{qu2024muptgenerativesymbolicmusic} introduced a scalable pretraining model using ABC notation and multi-track Transformer design. While powerful, Transformers still face challenges in extremely long-sequence generation due to error accumulation and quality degradation.

\subsubsection{Diffusion-based Methods}

Diffusion models offer a non-autoregressive generation approach with strong fidelity. Discrete diffusion models \citep{plasser2023discretediffusionprobabilisticmodels} have been applied to symbolic music generation, demonstrating state-of-the-art sample quality and flexible note-level infilling capabilities. Diff-Music \citep{nistal2024diffariffmusicalaccompanimentcocreation} first applied discrete diffusion to MIDI generation. Cascaded-Diff \citep{wang2024wholesonghierarchicalgenerationsymbolic} employed a multi-step diffusion sampling process to generate music progressively from high-level structure to detailed melody. However, diffusion models are computationally expensive for long sequences due to multi-step sampling, and often struggle with maintaining coherence and duration control.

\subsection{Hierarchical Music Generation}

Hierarchical music generation addresses long-sequence challenges by decomposing generation into multiple levels (e.g., sections, phrases, notes). Early models like Hierarchical RNN \citep{zhao2019hierarchicalrecurrentneuralnetwork} used multi-layer recurrent structures. SymphonyNet \citep{liu2022symphonygenerationpermutationinvariant} modeled movements, phrases, and notes for symphonic music. Cascaded-Diff \citep{wang2024wholesonghierarchicalgenerationsymbolic} integrated structural language modeling with diffusion techniques. Despite progress, current methods still face issues in inter-level information flow and precise duration control.

In summary, current symbolic music generation methods face challenges such as error accumulation, low efficiency, and structural inconsistency in long-sequence modeling. We propose an anchored cyclic generation paradigm and hierarchical framework that address these issues through novel generation mechanisms and architectural designs, thereby providing new solutions for long-sequence symbolic music generation.

\section{Method}

In this section, we introduce the Piano Token representation, the anchored cyclic generation paradigm, and the hierarchical anchored cyclic generation framework designed based on this paradigm for generating high-quality long-sequence symbolic music.

\begin{figure}[t]
\centering
\includegraphics[width=0.4\textwidth]{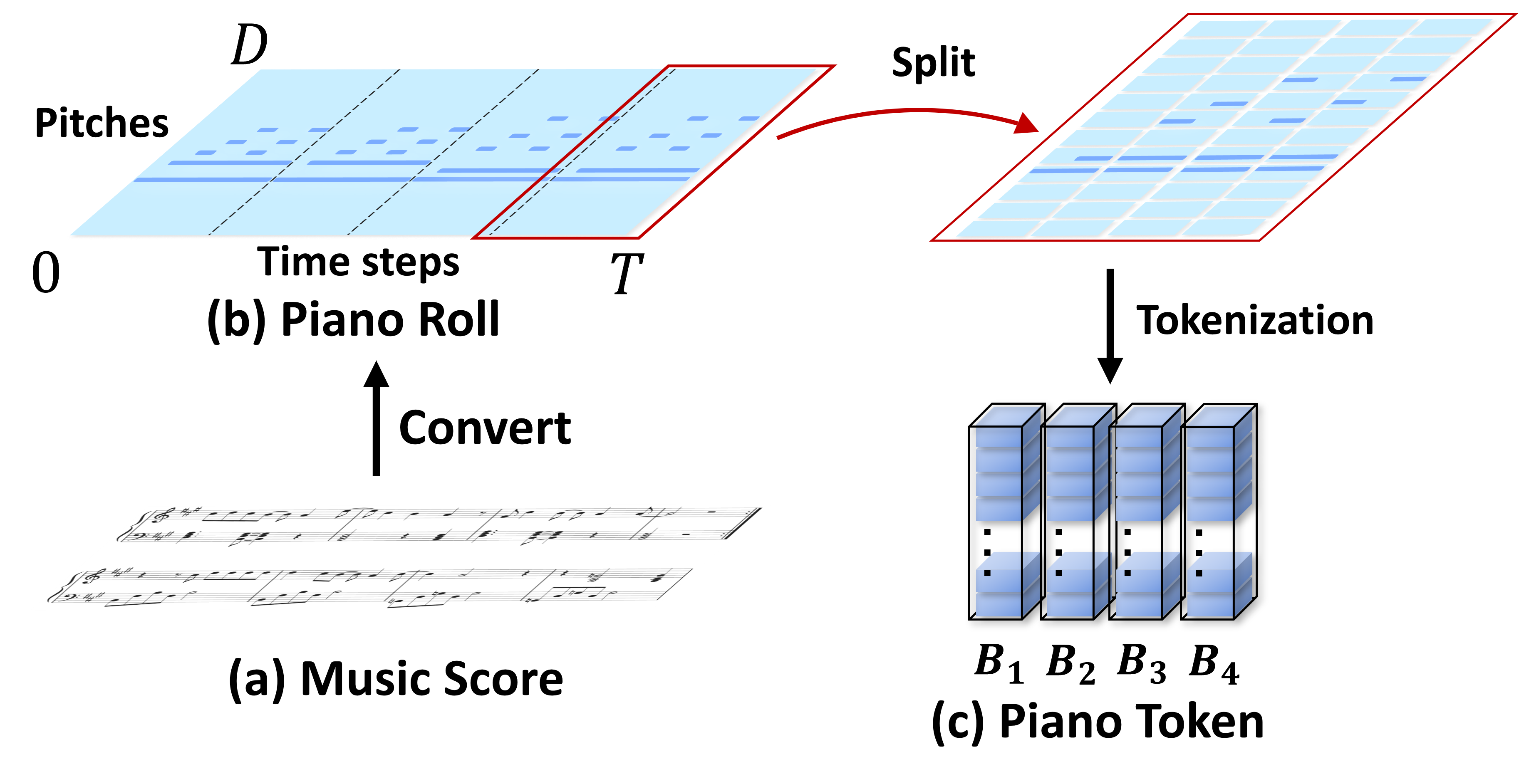}
\caption{Converting musical scores to piano tokens. (a) Example musical score. (b) Piano roll with $T$ time steps and $D$ pitch dimensions. The red-boxed section demonstrates split details. (c) Piano token representation, where each sub-unit of the split piano roll is tokenized into a piano token. Piano tokens in the same column collectively form a block.}
\label{fig1}
\end{figure}

\subsection{Piano Token Representation}

Common symbolic music representations primarily use MIDI event-based representations, such as REMI \citep{huang2020pop} or Compound Word representations \citep{hsiao2021compound}, or ABC notation representations. The sequence length of these representation forms exhibits a non-linear relationship with music duration. When music complexity is high and note changes are frequent, extremely long representation sequences are often required. This typically causes severe error accumulation and missing important tokens during music sequence generation.

To address this issue, we design the piano token representation based on piano roll, which is a more efficient music representation method shown in Figure \ref{fig1}. The piano token representation explicitly encodes temporal sequences, where the sequence length $L$ is positively correlated with music duration $T$, expressed as $L \propto T$. This representation method maps continuous music representations to sparse discrete representations through tokenization while preserving the original spatiotemporal structure. Specifically, we first split the piano roll representation into $N$ patches, where each patch contains $d \times t$ elements from the piano roll, with $d$ representing the number of pitch dimensions covered by a single patch and $t$ denoting the number of time steps covered by a single patch. We then tokenize each patch using a single token to encode its complete content, achieving data compression. Since the piano roll representation is a binary matrix of shape $D \times T$, where $D$ represents the pitch dimension and $T$ denotes the number of time steps. Each element $x_{p,t} \in \{0,1\}$ indicates whether pitch $p$ is activated at timestep $t$. The partitioned patches retain this characteristic, so the vocabulary size corresponding to the piano token representation is $2^{d*t}$. By adjusting the values of $d$ and $t$, we can flexibly control the patch size, thereby regulating the vocabulary scale and encoded sequence length to accommodate different application scenarios. After tokenization, the original matrix of size $D \times T$ is converted into a piano token matrix of shape $\frac{D}{d} \times \frac{T}{t}$. The piano token matrix representation serves as the core representational for music and participates in subsequent music generation processes.

We define each column of the piano token matrix as a block $B$, which contains musical information from $t$ consecutive time steps in the original piano roll representation. Each block contains complete musical fragments within short time intervals, and this block structure also plays a crucial role in subsequent music generation workflows.

\begin{figure*}[t]
\centering
\includegraphics[width=0.9\textwidth]{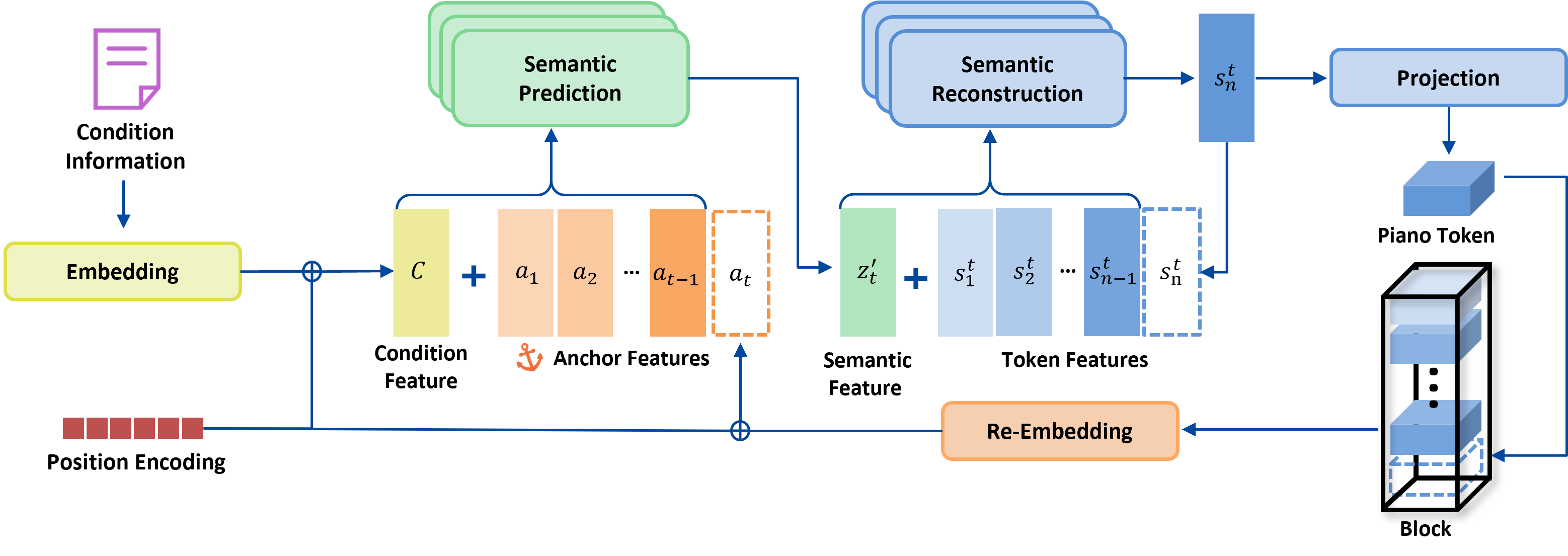}
\caption{ACG Paradigm Architecture. The embedding layer encodes the conditional information into feature vectors, concatenated with anchor features from existing content (anchor features are absent at the initial time step) and feeds into the semantic prediction model. The semantic prediction model generates semantic features $z$ for the current time step and feeds it to semantic reconstruction model, which autoregressively generate a sequence of features corresponding to piano tokens $s$ through multiple iterations. The generated token feature $s$ concatenates with semantic feature $z$ for iteration. The $n$ piano tokens combine into a block as final output for semantic feature $z$, then convert into an anchor feature for iterative generation by re-embedding model.}
\label{fig2}
\end{figure*}

\subsection{Anchored Cyclic Generation Paradigm}

Error accumulation is a common problem in autoregressive models, where discrepancies exist between the model's generation and optimal prediction value in each iteration round, and these errors accumulate throughout the iterative process, ultimately leading to severe degradation of the overall generation quality. Error accumulation can be regarded as the primary factor contributing to the degradation of generation quality in long-sequence music generation. To mitigate this issue, we propose the ACG, a novel generation paradigm that can significantly reduce error accumulation during long sequence generation. We draw inspiration from teacher forcing training methodology: when generating symbolic music sequences, at each iteration, the model predicts features for the next time step $t$ based on determined historical information, which we call anchor features $A_{t-1} = \{a_1, a_2, a_3, ..., a_{t-1}\}$, thereby minimizing the error between the current prediction and the optimal value.

As illustrated in Figure \ref{fig2}, an ACG structure comprises three key components: a semantic prediction model, a semantic reconstruction model, and a specialized re-embedding layer. The semantic prediction model and semantic reconstruction model consist of two cascaded transformer decoder models. The semantic prediction model is responsible for predicting the semantic features $z_t'$ of the current time step $t$ based on input conditions $C$ and anchor features $A_{t-1}$. It predicts the content of a whole block, which contains token combinations' information. The expression is as follows:

\[
z_t' = f_{\text{sem}}(A_{t-1}, C, t; \theta_{\text{sem}})
\]

\noindent The semantic reconstruction model decodes the semantic feature $z_t'$ and projects it into Piano Token sequence $S_{\text{token}}^{t} = \{s_1^{t}, s_2^{t}, s_3^{t}, ..., s_n^{t}\}$ via an additional linear layer, where $n$ denotes the sequence length. The semantic reconstruction process can be expressed as follows:

\[
S_{\text{token}}^{t} = f_{\text{proj}}(f_{\text{rec}}(z_t'; \theta_{\text{rec}}); W_{\text{proj}}, b_{\text{proj}})
\]

\noindent The re-embedding layer is a neural network composed of multiple fully connected layers, which is responsible for remapping the reconstructed token sequence $S_{\text{token}}^{t}$ back to anchor features $a_t$ for generation in the next iteration.

\[
a_t = f_{\text{reemb}}(B_t; \theta_{\text{reemb}})
\]

\begin{figure}[t]
\centering
\includegraphics[width=0.9\columnwidth]{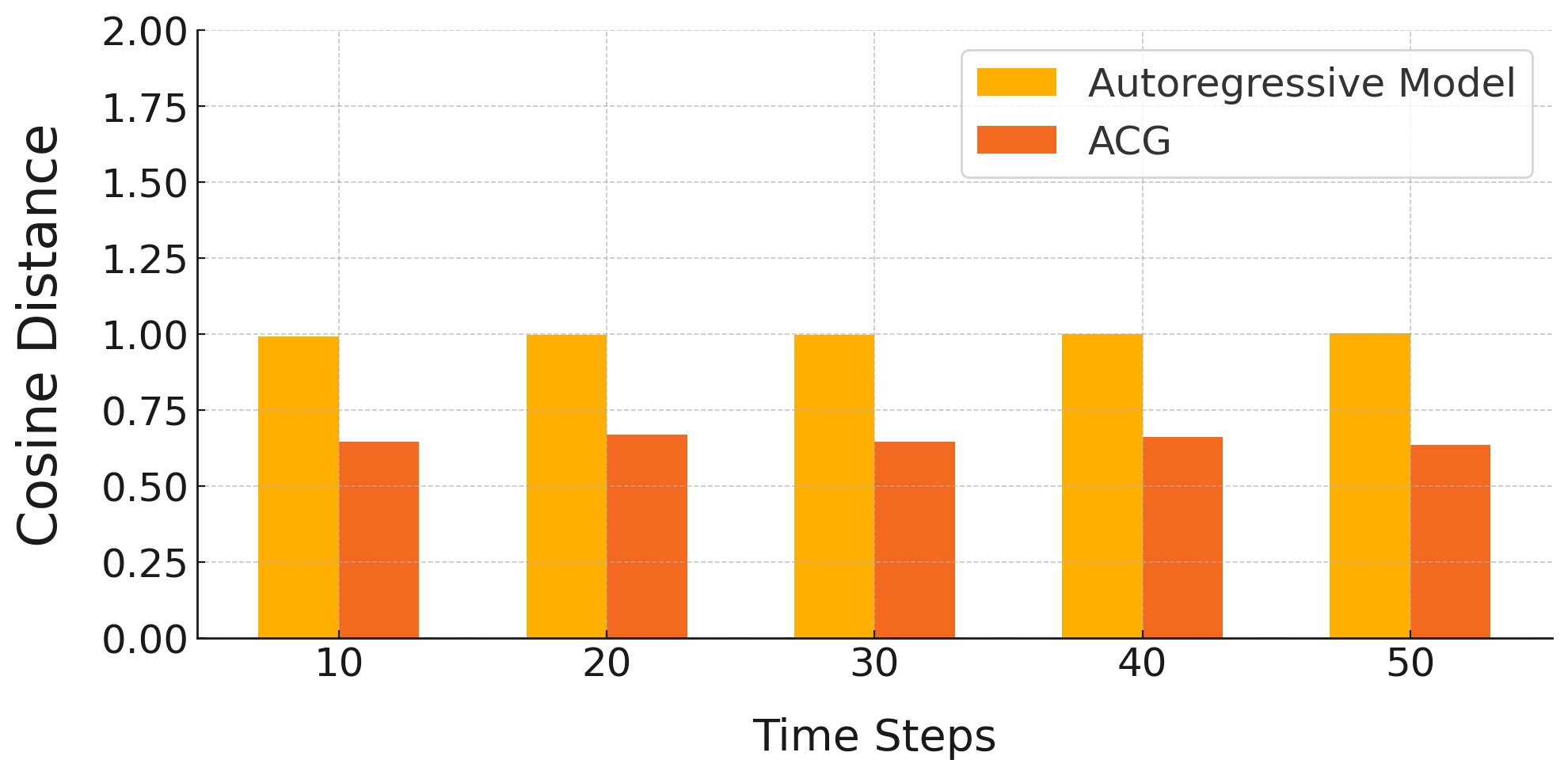}
\caption{Cosine distances between predicted and ground truth feature vectors for the ACG paradigm and conventional autoregressive models across iterations. The ACG paradigm consistently achieves lower cosine distances compared to conventional autoregressive models.}
\label{fig3}
\end{figure}

In our design, the three components of the ACG paradigm are jointly trained in an end-to-end manner. It should be noted that in the ACG paradigm, the semantic prediction model does not independently generate all latent vectors $A$ of semantic information before the semantic restoration model sequentially restores them to token representations. In our method, the semantic prediction model first predicts the semantic latent feature $z_t'$ for the current time step $t$ and transmits it to the semantic restoration model and additional projection layer, which then autoregressively decodes a sequence $S_{\text{token}}^{t}$ containing all tokens in the block based on feature $z_t'$. Each element in sequence $S_{\text{token}}^{t}$ is stacked and rearranged to form the final output block $B_t$ for the current time step. Additionally, we treat the block $B_t$ obtained in each iteration as confirmed historical information and input $B_t$ into a re-embedding layer to transform it back into an anchor semantic feature $a_t$. The anchor feature $a_t$ for the current time step is concatenated with features from all previous time steps $A_{t-1}$ and fed into the semantic decoder for semantic feature prediction of the next time step. This anchor feature derived from confirmed content can better approximate the optimal feature, guiding the semantic prediction model to achieve more accurate outputs through what we refer to as the anchor mechanism. In the task of generating subsequent musical content given an opening, we extracted 100 samples each of semantic features $z'$ from the ACG paradigm and semantic features $z$ from conventional autoregressive methods, and compared them by computing cosine distances with ground truth, thereby confirming that our hypothesis holds in practice. The result shown in Figure \ref{fig3}. Compared to traditional autoregressive models, the ACG paradigm achieves an average reduction of 34.7\% in cosine distance between predicted feature vectors and ground-truth semantic vectors. For the mathematical proof of the effectiveness of the ACG paradigm, please refer to the supplementary materials.

This demonstrates that our proposed ACG paradigm effectively mitigates the error accumulation inherent in autoregressive models, thereby achieving superior generation performance. In terms of time complexity, ACG also outperforms conventional approaches. The time complexity $O(L^2)$ of conventional autoregressive models scales quadratically with increasing sequence length $L$, whereas the time complexity of the ACG paradigm is $O(L_{\text{sem}}^2) + L_{\text{sem}} \times O(L_{\text{rec}}^2)$, where $L_{\text{sem}}$ represents the length of semantic features sequence $Z'$, and $L_{\text{rec}}$ represents the length of token sequence $S$. The ACG paradigm decomposes autoregressive generation tasks into a two-stage subtask framework, implemented through employing separate semantic prediction and semantic restoration models. The semantic prediction model operates solely at the block level to predict semantic features, while the semantic restoration model focuses exclusively on reconstructing fixed-length token sequences from block features. This task decomposition significantly reduces the computational burden of models in long-sequence autoregressive generation tasks, with the efficiency gains becoming more pronounced as sequence length increases.

\begin{figure*}[t]
\centering
\includegraphics[width=0.9\textwidth]{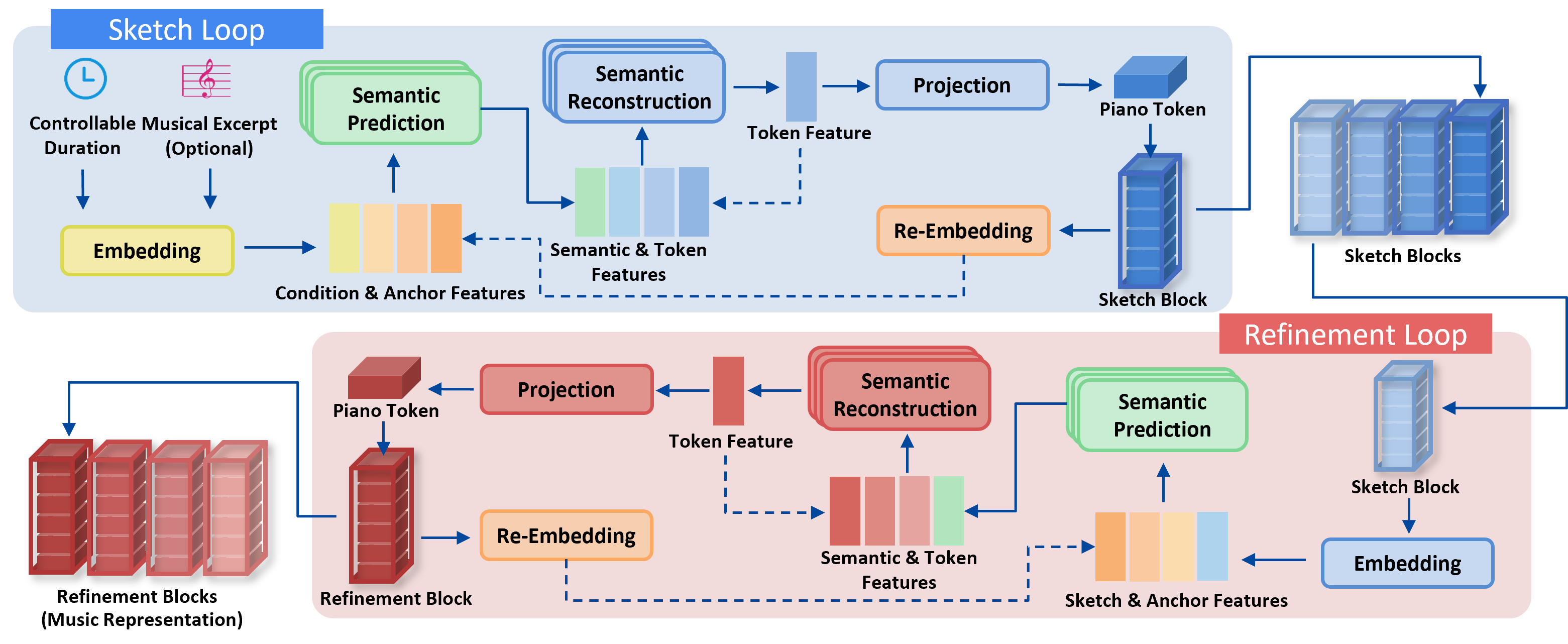}
\caption{The Hi-ACG framework, which comprises a sketch loop and a refinement loop. The sketch loop takes duration condition and optional musical constraints as input, generating a musical sketch with coarse-grained information for the entire composition using block-by-block processing. The refinement loop then sequentially processes the sketch blocks, transforming each into multiple detailed blocks with rich musical representations.}
\label{fig4}
\end{figure*}

\subsection{Hierarchical Anchored Cyclic Generation Framework}

We propose the Hi-ACG framework, built upon the ACG paradigm, to generate high-quality, long-sequence symbolic music with complete structure and precise duration control. This cascaded model architecture embodies a core principle: simulating human compositional cognition through hierarchical music generation that proceeds from global structure to local refinement. This approach mirrors how composers naturally work---first establishing the overall structural framework, then gradually developing specific musical details. By doing so, the framework maintains both global musical coherence and rich local expression, ultimately producing more natural and musically acceptable compositions.

The Hi-ACG framework features two interconnected loops: the Sketch Loop and the Refinement Loop. The Sketch Loop generates high-level structural sketches that establish the compositional backbone. Building on this sketch information, the Refinement Loop focuses on creating rich expressive details, transforming abstract structures into concrete musical content. This clear division of responsibilities allows the framework to optimize music generation at different levels of abstraction, ensuring both structural coherence in long sequences and precise control over duration while enhancing local musical quality.

We propose an approach where the Sketch Loop and Refinement Loop are trained separately with different objectives. We obtain training data for the Sketch Loop by resampling real music data, performing two samplings within each measure to extract each measure's core musical information. This approach preserves the main musical characteristics of each measure while significantly reducing sequence length. After resampling the entire composition, we convert the results to piano token representation for Sketch Loop training. The Refinement Loop uses paired training, utilizing sketches generated by the Sketch Loop paired with corresponding complete musical pieces. We also convert the training data to piano token representation to maintain consistency. During training, the Refinement Loop learns to expand sketch information into complete, detailed musical content, learning to map abstract structures to concrete musical expressions.

During the music generation phase, the framework follows this workflow: First, the Sketch Loop generates a piano token matrix of the complete composition sketch based on conditions such as duration or musical input, providing the overall structural framework. The sketch is then segmented into sequence $B_{\text{sketch}} = \{b_1, b_2, ..., b_t\}$ block-by-block, with each element in sequence $B_{\text{sketch}}$ passed individually to the Refinement Loop for processing. The Refinement Loop uses each input block $b_t$ to expand and generate corresponding detailed musical content $B_{\text{refinement}}$. Finally, all refined block sequences are combined in chronological order to form the complete musical work. Through this hierarchical generation strategy, our framework can generate high-quality long-sequence symbolic music with rich detailed expression while ensuring overall structural coherence. For structural control, the global planning of the Sketch Loop ensures that generated music maintains coherent overall structure, avoiding the structural drift commonly seen in long-sequence generation. To ensure quality, the Refinement Loop focuses on optimizing local musical expression, producing music that maintains structural coherence while featuring rich musical details and expressiveness. Moreover, the hierarchical design provides the framework with strong scalability, enabling it to handle music sequences of arbitrary length and making it technically feasible to generate ultra-long musical compositions.

\section{Experiment}

To validate the Hi-ACG framework's effectiveness, we conduct comprehensive objective and subjective experiments evaluating the generation quality. This chapter presents the experimental setup and evaluation results.

\subsection{Experimental Setup}

\subsubsection{Dataset}

We train our model on the MuseScore dataset and the POP909 dataset. The MuseScore dataset contains 140,000 two-track piano scores lasting 1-3 minutes. We convert them to piano rolls in multi-hot array format, with a minimum resolution of 1/16 beat. Each time step preserves 88 pitches corresponding to standard piano keys. We encode the piano rolls into piano tokens to train our model. For POP909, we similarly convert them into piano roll representations with the same temporal resolution, then transform them into piano tokens.

\subsubsection{Details}

During data preprocessing in our experiments, we set $d$ to 2 and $t$ to 4, with each token corresponding to a patch range of $2 \times 4$ elements. This results in a piano tokens matrix of size $44 \times (\frac{T}{4})$. For model hyperparameters, in the ACG paradigm, the semantic prediction model contains 12 self-attention layers, the semantic reconstruction model contains 6 self-attention layers, and the re-embedding layer consists of 3 fully connected layers, all models use a hidden dimension of 1024. We trained on the MuseScore data for 30 epochs using 4 NVIDIA RTX 4090 GPUs, then continued fine-tuning our pre-trained model using POP909 data to improve generation performance.

\subsection{Evaluation}

\subsubsection{Baseline}

We compare against Transformer-based models (Music Transformer, BPE Transformer) and diffusion-based model (Cascaded-Diff), all fine-tuned on the same datasets for fair comparison. We also conduct ablation studies on Hi-ACG components. In tables, ``GT" denotes ground truth, ``MT" denotes Music Transformer, ``BT" denotes BPE Transformer, ``CD" denotes Cascaded-Diff, ``Full" denotes the complete framework, ``SL" denotes the sketch loop, and ``SP" denotes the semantic prediction.

\subsubsection{Objective Evaluation}

To objectively evaluate of the music generated by various models, we design specialized music evaluation metrics. The evaluation metrics encompass four aspects, assessing both model-generated and real music from pitch, rhythm, harmony, and melody. Pitch evaluation employs information entropy to quantify note diversity within a musical piece. The information entropy is calculated as follows, where $p(i)$ represents the probability of note occurrence. Higher pitch entropy indicates greater pitch diversity with a more uniform distribution.

\[
H_{\text{pitch}} = -\sum_{i=1}^{n} p(i) \log_2 p(i)
\]

\noindent Similar to pitch evaluation, rhythm evaluation employs entropy to measure rhythmic complexity by analyzing the frequency and distribution of various note durations. Here, $p(j)$ denotes the probability of each duration type.

\[
H_{\text{rhythm}} = -\sum_{j=1}^{n} p(j) \log_2 p(j)
\]

\noindent Harmonic consistency evaluates the degree of matching between notes and tonality. We identify the musical tonality using Music21's \citep{cuthbert2010music21} key analysis algorithm, then calculate the proportion of notes that belong to the tonal distribution. The melodic smoothness metric is based on melodic fluency principles in music theory, assessing smoothness by analyzing the size of intervals between adjacent notes in the melody. Specifically, we define intervals exceeding a perfect fourth as large leaps and calculate the proportion of large leaps in the melody, where more frequent large leaps reduce the perceived musical quality. Additionally, we employ an LLM for music quality assessment. We use the Qwen3-235B-A22B model to evaluate musical quality, providing comprehensive scores considering multiple dimensions including melody, rhythm, and arrangement. The scoring employs the MOS (Mean Opinion Score) method with a score range of 1-5, where 1 is the lowest and 5 is the highest. We conduct objective evaluation across three tasks: 30-second short music generation, 2-minute long music generation, and conditional input long music generation. Short music generation evaluation represents local music quality, while long music generation assesses fluency, structure, and compositional completeness. Conditional music generation show the model's ability to continue music from given input, demonstrating music understanding and completion capabilities.

\begin{table}[t]
\centering
\small
\setlength{\tabcolsep}{1mm}
\begin{tabular}{llllll}
\hline
& Pitch & Rhythm & Harmony & Melody & LLM score \\
\hline
GT & 1.92 & 1.43 & 0.87 & 0.52 & 3.50 \\
\hline
MT & \textbf{1.95} & \textbf{1.66} & 0.94 & 0.41 & 2.25 \\
BT & 3.16 & 1.74 & 0.90 & \textbf{0.55} & 2.43 \\
CD & 3.26 & 2.36 & 0.91 & 0.66 & \textbf{3.37} \\
\hline
w/o SL \& SP & 2.44 & 1.80 & 0.94 & 0.40 & 2.22 \\
w/o SL & 1.32 & 1.71 & 0.84 & 0.62 & 3.06 \\
Full & 1.43 & 1.69 & \textbf{0.89} & 0.60 & 3.10 \\
\hline
\end{tabular}
\caption{Objective evaluation results for 30-seconds unconditional music generation. Performance improves when Pitch, Rhythm, Harmony, and Melody values match the ground truth. Higher LLM scores show better performance.}
\label{table1}
\end{table}

\begin{table}[t]
\centering
\small
\setlength{\tabcolsep}{1mm}
\begin{tabular}{llllll}
\hline
& Pitch & Rhythm & Harmony & Melody & LLM score \\
\hline
GT & 2.20 & 1.06 & 0.90 & 0.50 & 3.55 \\
\hline
MT & 3.49 & 3.19 & 0.92 & 0.29 & 2.29 \\
BT & 3.16 & 1.74 & \textbf{0.90} & 0.55 & 2.45 \\
CD & 3.38 & 2.30 & 0.91 & 0.90 & 2.89 \\
\hline
w/o SL \& SP & - & - & - & - & - \\
w/o SL & 1.56 & 0.84 & 0.83 & 0.41 & 2.92 \\
Full & \textbf{2.43} & \textbf{1.03} & \textbf{0.90} & \textbf{0.47} & \textbf{3.17} \\
\hline
\end{tabular}
\caption{Objective evaluation results for 2-minutes unconditional music generation. Performance improves when Pitch, Rhythm, Harmony, and Melody values match the ground truth. Higher LLM scores show better performance.}
\label{table2}
\end{table}

\begin{table}[t]
\centering
\small
\setlength{\tabcolsep}{1mm}
\begin{tabular}{llllll}
\hline
& Pitch & Rhythm & Harmony & Melody & LLM score \\
\hline
GT & 2.20 & 1.06 & 0.90 & 0.50 & 3.55 \\
\hline
MT & 3.73 & 3.53 & 0.96 & 0.37 & 2.24 \\
BT & 3.89 & 3.20 & 0.88 & 0.66 & 2.33 \\
CD & - & - & - & - & - \\
\hline
w/o SL \& SP & - & - & - & - & - \\
w/o SL & 2.69 & 1.90 & 0.99 & 0.33 & 3.05 \\
Full & \textbf{2.19} & \textbf{1.27} & \textbf{0.91} & \textbf{0.43} & \textbf{3.30} \\
\hline
\end{tabular}
\caption{Objective evaluation results for 2-minutes conditional music generation. Performance improves when Pitch, Rhythm, Harmony, and Melody values match the ground truth. Higher LLM scores show better performance.}
\label{table3}
\end{table}

\subsubsection{Subjective Evaluation}

In music generation tasks, subjective evaluation often provides a more intuitive reflection of the quality of the generated music's impact on listeners. We also conducted a comparison between our model and the baseline in subjective evaluation. The subjective evaluation employed the MOS method, where evaluators rate each music sample on a scale from 1 to 5, with higher scores indicating superior musical quality. Evaluators were blinded to which model generated each music sample. The evaluators were 79 volunteers with music education backgrounds, including 46 males and 33 females.

\begin{table}[t]
\centering
\small
\setlength{\tabcolsep}{1mm}
\begin{tabular}{cccccccc}
\hline
& GT & MT & BT & CD & w/o SL \& SP & w/o SL & Full \\
\hline
Score & 3.31 & 1.96 & 2.05 & 2.91 & 1.85 & 2.52 & \textbf{3.02} \\
\hline
\end{tabular}
\caption{Subjective evaluation for 2-minutes music generation use MOS evaluation, higher score indicate superior performance.}
\label{table4}
\end{table}

\section{Conclusion}

Long-sequence symbolic music generation faces a critical challenge: error accumulation that degrades musical structure and fluency. To address this, we introduce the ACG paradigm with piano token representation and propose a hierarchical music generation framework. This approach enhances generation quality while enabling flexible conditional music generation. Our experiments demonstrate substantial improvements in both statistical metrics and overall music quality, with both objective metrics and subjective evaluations consistently validating our approach's superiority. The ACG paradigm represents a major breakthrough in long-sequence symbolic music generation, providing a flexible framework easily integrated into existing autoregressive models. The system adapts well to downstream tasks through fine-tuning and offers valuable insights for music understanding applications. These foundational principles pave the way for more sophisticated and controllable symbolic music generation systems. While demonstrated on symbolic music, the core principle of ACG---using confirmed content as anchors to mitigate error accumulation---is broadly applicable to other long-sequence generation tasks, including structured text generation and hierarchical content synthesis, opening promising directions for future research.

\section*{Limitations}

Our method currently lacks fine-grained control during generation, limiting dynamic adjustment for personalized music creation. Future work will integrate additional tokens capturing musical expression and structural elements to improve controllability. Additionally, while the piano token representation achieves efficient compression, it may lose subtle timing nuances present in event-based representations. The current framework focuses on piano music; extending to diverse instruments and timbres requires further investigation. We also plan to extend ACG to multi-track symbolic music generation and explore its application to other sequential generation domains.

\bibliography{references}

\appendix

\section{Piano Token Conversion Algorithm}
\label{sec:appendix_piano_token}

This section provides the conversion algorithms for the piano token representation. Algorithm~\ref{alg:pianoroll_to_tokens} describes how to convert a piano roll representation into piano tokens, and Algorithm~\ref{alg:tokens_to_pianoroll} shows the reverse conversion.

\begin{algorithm}[t]
\caption{Piano roll to piano tokens conversion}
\label{alg:pianoroll_to_tokens}
\small
\begin{algorithmic}[1]
\Require Piano roll matrix $\mathbf{P} \in \{0,1\}^{88 \times W}$
\Ensure Token matrix $\mathbf{S} \in \{0,1,\ldots,255\}^{W/4 \times 44}$
\State $h_p \gets 2, w_p \gets 4$ \Comment{Patch dimensions}
\State $N_r \gets 88 / h_p = 44$, $N_c \gets W / w_p$
\State Initialize $\mathbf{S} \in \mathbb{Z}^{N_c \times N_r}$
\For{$j = 0$ \textbf{to} $N_c - 1$}
    \For{$i = 0$ \textbf{to} $N_r - 1$}
        \State $\mathbf{patch} \gets \mathbf{P}[i \cdot h_p:(i+1) \cdot h_p, j \cdot w_p:(j+1) \cdot w_p]$
        \State $\mathbf{flat} \gets \text{flatten}(\mathbf{patch})$
        \State $token \gets 0$
        \For{$k = 0$ \textbf{to} $7$}
            \State $token \gets (token \ll 1) \lor \mathbf{flat}[k]$
        \EndFor
        \State $\mathbf{S}[j, i] \gets token$
    \EndFor
\EndFor
\State \Return $\mathbf{S}$
\end{algorithmic}
\end{algorithm}

\begin{algorithm}[t]
\caption{Piano tokens to piano roll conversion}
\label{alg:tokens_to_pianoroll}
\small
\begin{algorithmic}[1]
\Require Token matrix $\mathbf{S} \in \{0,1,\ldots,255\}^{N_c \times N_r}$
\Ensure Reconstructed piano roll $\mathbf{P'} \in \{0,1\}^{88 \times 4N_c}$
\State $h_p \gets 2, w_p \gets 4$
\State Initialize $\mathbf{P'} \in \{0,1\}^{N_r \cdot h_p \times N_c \cdot w_p}$
\For{$j = 0$ \textbf{to} $N_c - 1$}
    \For{$i = 0$ \textbf{to} $N_r - 1$}
        \State $token \gets \mathbf{S}[j, i]$
        \State $\mathbf{binary} \gets \text{int2bin}(token, 8)$
        \State $\mathbf{patch} \gets \text{reshape}(\mathbf{binary}, (h_p, w_p))$
        \State $\mathbf{P'}[i \cdot h_p:(i+1) \cdot h_p, j \cdot w_p:(j+1) \cdot w_p] \gets \mathbf{patch}$
    \EndFor
\EndFor
\State \Return $\mathbf{P'}$
\end{algorithmic}
\end{algorithm}

\section{Mathematical Proof of ACG Effectiveness}
\label{sec:appendix_proof}

We provide mathematical derivations validating the ACG paradigm's effectiveness in reducing error accumulation.

Let $\varepsilon_t$ denote the local error at step $t$ for conventional autoregressive models, and $\varepsilon'_t$ for ACG. The cumulative errors are:
\[
\delta_n = \sum_{t=1}^{N} \varepsilon_t, \quad \delta'_n = \sum_{i=1}^{N} \varepsilon'_i
\]

We assume that features $z_t'$ predicted based on anchor features $A_{t-1}$ are closer to optimal features $h_t$ than purely autoregressive predictions $z_t$:
\[
\forall t,\ \|z_t' - h_{t}\| \leq \|z_t - h_{t}\|
\]

Since $\varepsilon_t = z_t - h_{t}$ and $\varepsilon'_t = z_t' - h_{t}$, by the anchoring mechanism, $\varepsilon'_t \leq \varepsilon_t$. Using MSE loss:
\[
L_1 = \frac{1}{N}\sum_{t=1}^{N}(z_t - h_t)^2, \quad L_2 = \frac{1}{N}\sum_{t=1}^{N}(z'_t - h_t)^2
\]

Since the loss function satisfies the Lipschitz condition $|f(x) - f(y)| \leq K \|x - y\|$:
\[
|L_2 - L_1| \leq K \left\| \sum_{t=1}^{N} (\varepsilon_t' - \varepsilon_t) \right\| \leq 0
\]

This demonstrates that ACG effectively mitigates error accumulation.

\section{Time Complexity Analysis}
\label{sec:appendix_complexity}

Conventional autoregressive models have time complexity $\mathcal{O}(L^2 \cdot d)$ where $L$ is sequence length and $d$ is hidden dimension.

The ACG paradigm decomposes this into two stages:
\begin{itemize}
\item \textbf{Semantic Prediction}: $\mathcal{C}_{\text{sem}} = \mathcal{O}(L_{sem}^2 \cdot d)$
\item \textbf{Semantic Reconstruction}: $\mathcal{C}_{\text{rec}} = L_{sem} \times \mathcal{O}(L_{rec}^2 \cdot d)$
\end{itemize}

Total ACG complexity:
\[
\mathcal{C}_{\text{ACG}} = \mathcal{O}(L_{sem}^2 \cdot d) + L_{sem} \times \mathcal{O}(L_{rec}^2 \cdot d)
\]

Given $L = L_{sem} \times L_{rec}$ and fixed $L_{rec}$, the speedup factor is $\mathcal{O}(L_{rec}^2)$, making ACG particularly effective for long sequences.

\section{Extended Cosine Distance Comparison}
\label{sec:appendix_cosine}

\begin{figure}[t]
\centering
\includegraphics[width=0.9\columnwidth]{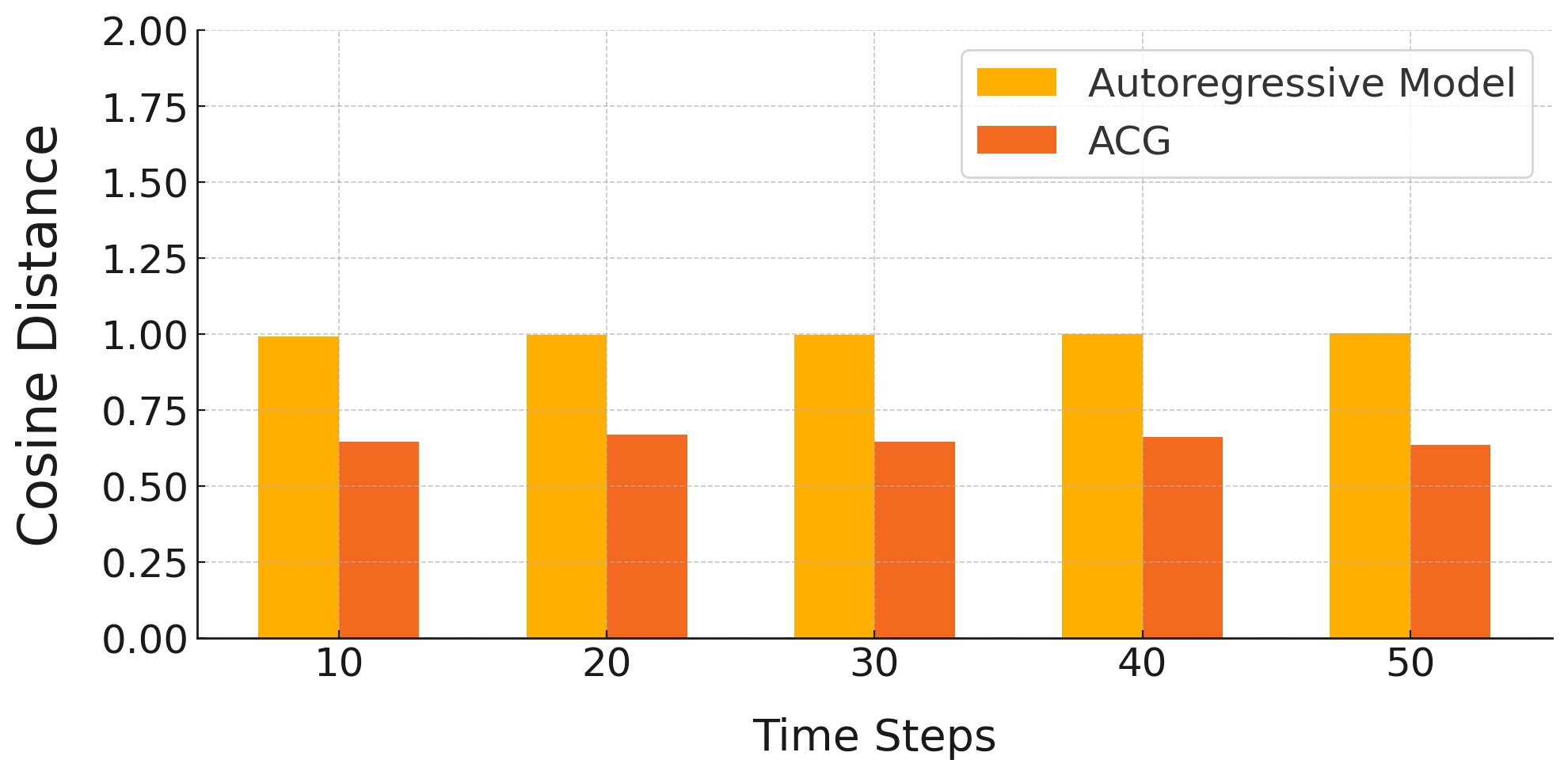}
\caption{Cosine distance comparison across 50 iterative steps.}
\label{fig:cosine_50}
\end{figure}

\begin{figure}[t]
\centering
\includegraphics[width=0.9\columnwidth]{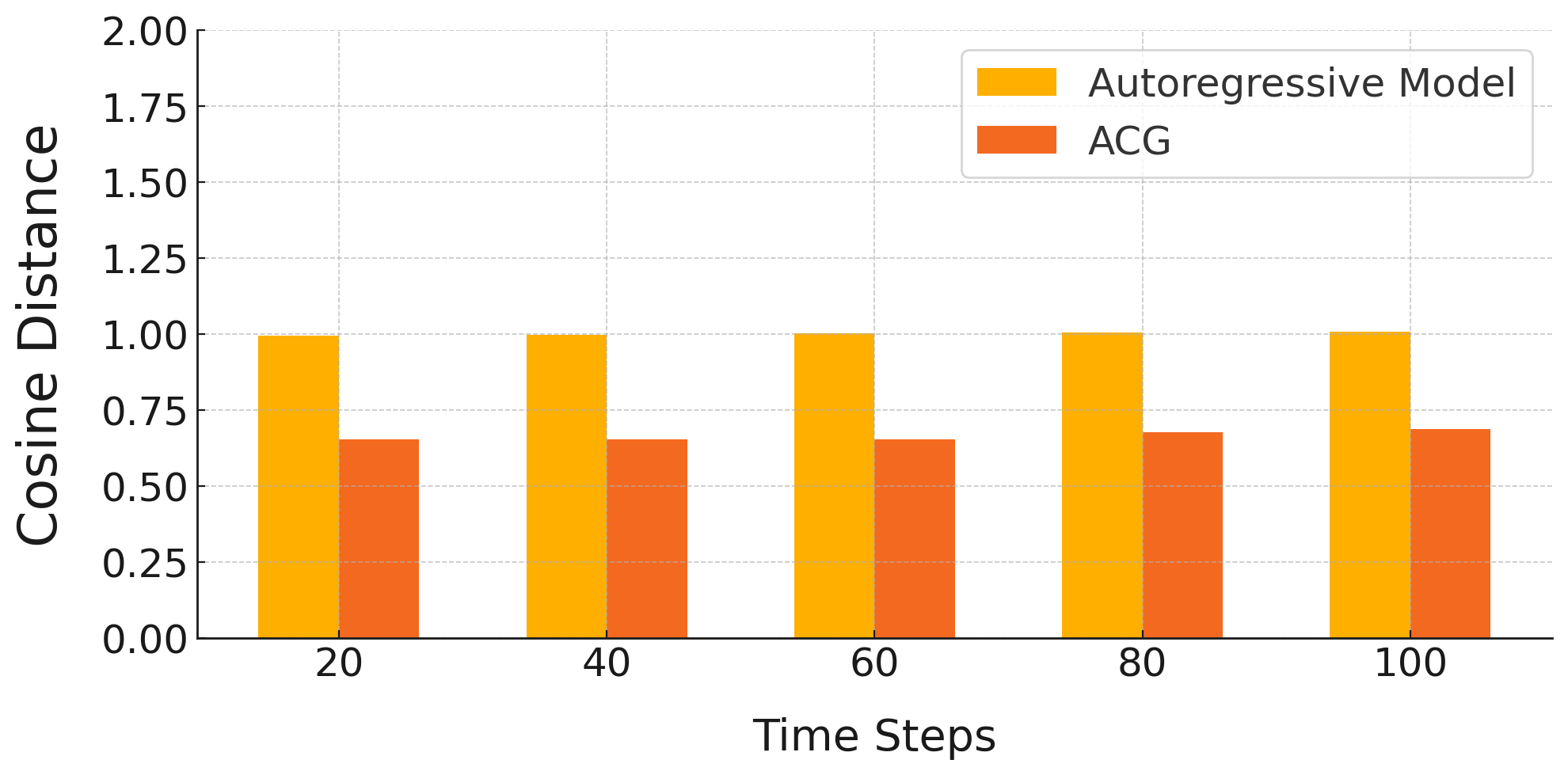}
\caption{Cosine distance comparison across 100 iterative steps.}
\label{fig:cosine_100}
\end{figure}

Figure~\ref{fig:cosine_50} and Figure~\ref{fig:cosine_100} show cosine distances between predicted and ground truth features for 50 and 100 timesteps respectively, demonstrating ACG's consistent advantage.

\section{Ablation Studies}
\label{sec:appendix_ablation}

\subsection{Architectural Ablation}

We investigate the effects of model size with Small (4+8 layers), Middle (5+10 layers), and Large (6+12 layers) configurations. Results in Tables~\ref{table:arch_30s},~\ref{table:arch_2m}, and~\ref{table:arch_2mc} show the Large model achieves best performance.

\begin{table}[t]
\centering
\small
\setlength{\tabcolsep}{1mm}
\begin{tabular}{llllll}
\hline
& Pitch & Rhythm & Harmony & Melody & LLM \\
\hline
GT & 1.92 & 1.43 & 0.87 & 0.52 & 3.50 \\
\hline
Small & 1.40 & 1.75 & 0.77 & 0.40 & 3.01 \\
Middle & 1.41 & \textbf{1.68} & 0.84 & \textbf{0.59} & 3.08 \\
Large & \textbf{1.43} & 1.69 & \textbf{0.89} & 0.60 & \textbf{3.10} \\
\hline
\end{tabular}
\caption{Architectural ablation on 30-second generation.}
\label{table:arch_30s}
\end{table}

\begin{table}[t]
\centering
\small
\setlength{\tabcolsep}{1mm}
\begin{tabular}{llllll}
\hline
& Pitch & Rhythm & Harmony & Melody & LLM \\
\hline
GT & 2.20 & 1.06 & 0.90 & 0.50 & 3.55 \\
\hline
Small & 2.45 & \textbf{1.08} & 0.94 & 0.55 & 3.11 \\
Middle & 2.51 & 1.03 & 0.92 & 0.46 & 3.12 \\
Large & \textbf{2.43} & 1.03 & \textbf{0.90} & \textbf{0.47} & \textbf{3.17} \\
\hline
\end{tabular}
\caption{Architectural ablation on 2-minute unconditional generation.}
\label{table:arch_2m}
\end{table}

\begin{table}[t]
\centering
\small
\setlength{\tabcolsep}{1mm}
\begin{tabular}{llllll}
\hline
& Pitch & Rhythm & Harmony & Melody & LLM \\
\hline
GT & 2.20 & 1.06 & 0.90 & 0.50 & 3.55 \\
\hline
Small & 2.16 & 1.33 & 0.82 & 0.40 & 3.25 \\
Middle & 2.15 & 1.35 & 0.87 & 0.42 & 3.28 \\
Large & \textbf{2.19} & \textbf{1.27} & \textbf{0.91} & \textbf{0.43} & \textbf{3.30} \\
\hline
\end{tabular}
\caption{Architectural ablation on 2-minute conditional generation.}
\label{table:arch_2mc}
\end{table}

\subsection{Hyperparameter Ablation}

We investigate patch dimensions $(d, t)$ affecting vocabulary size and sequence lengths. Tables~\ref{table:hyper_30s},~\ref{table:hyper_2m}, and~\ref{table:hyper_2mc} show Patch$_{2,4}$ achieves optimal balance.

\begin{table}[t]
\centering
\small
\setlength{\tabcolsep}{1mm}
\begin{tabular}{llllll}
\hline
& Pitch & Rhythm & Harmony & Melody & LLM \\
\hline
GT & 1.92 & 1.43 & 0.87 & 0.52 & 3.50 \\
\hline
Patch$_{1,4}$ & 2.45 & 1.88 & 0.77 & 0.42 & 2.89 \\
Patch$_{2,4}$ & \textbf{1.43} & 1.69 & \textbf{0.89} & 0.60 & \textbf{3.10} \\
Patch$_{3,4}$ & 2.53 & \textbf{1.51} & 0.82 & \textbf{0.55} & 3.06 \\
\hline
\end{tabular}
\caption{Hyperparameter ablation on 30-second generation.}
\label{table:hyper_30s}
\end{table}

\begin{table}[t]
\centering
\small
\setlength{\tabcolsep}{1mm}
\begin{tabular}{llllll}
\hline
& Pitch & Rhythm & Harmony & Melody & LLM \\
\hline
GT & 2.20 & 1.06 & 0.90 & 0.50 & 3.55 \\
\hline
Patch$_{1,4}$ & 2.81 & 1.22 & 0.95 & \textbf{0.53} & 3.02 \\
Patch$_{2,4}$ & \textbf{2.43} & 1.03 & \textbf{0.90} & 0.47 & \textbf{3.17} \\
Patch$_{3,4}$ & 2.45 & \textbf{1.02} & 0.91 & 0.45 & 3.14 \\
\hline
\end{tabular}
\caption{Hyperparameter ablation on 2-minute unconditional generation.}
\label{table:hyper_2m}
\end{table}

\begin{table}[t]
\centering
\small
\setlength{\tabcolsep}{1mm}
\begin{tabular}{llllll}
\hline
& Pitch & Rhythm & Harmony & Melody & LLM \\
\hline
GT & 2.20 & 1.06 & 0.90 & 0.50 & 3.55 \\
\hline
Patch$_{1,4}$ & 2.24 & 1.34 & 0.95 & 0.56 & 3.19 \\
Patch$_{2,4}$ & \textbf{2.19} & \textbf{1.27} & 0.91 & 0.43 & \textbf{3.30} \\
Patch$_{3,4}$ & 2.18 & 1.20 & \textbf{0.90} & \textbf{0.49} & 3.22 \\
\hline
\end{tabular}
\caption{Hyperparameter ablation on 2-minute conditional generation.}
\label{table:hyper_2mc}
\end{table}

\section{Generation Examples}
\label{sec:appendix_examples}

\begin{figure*}[t]
\centering
\includegraphics[width=0.9\textwidth]{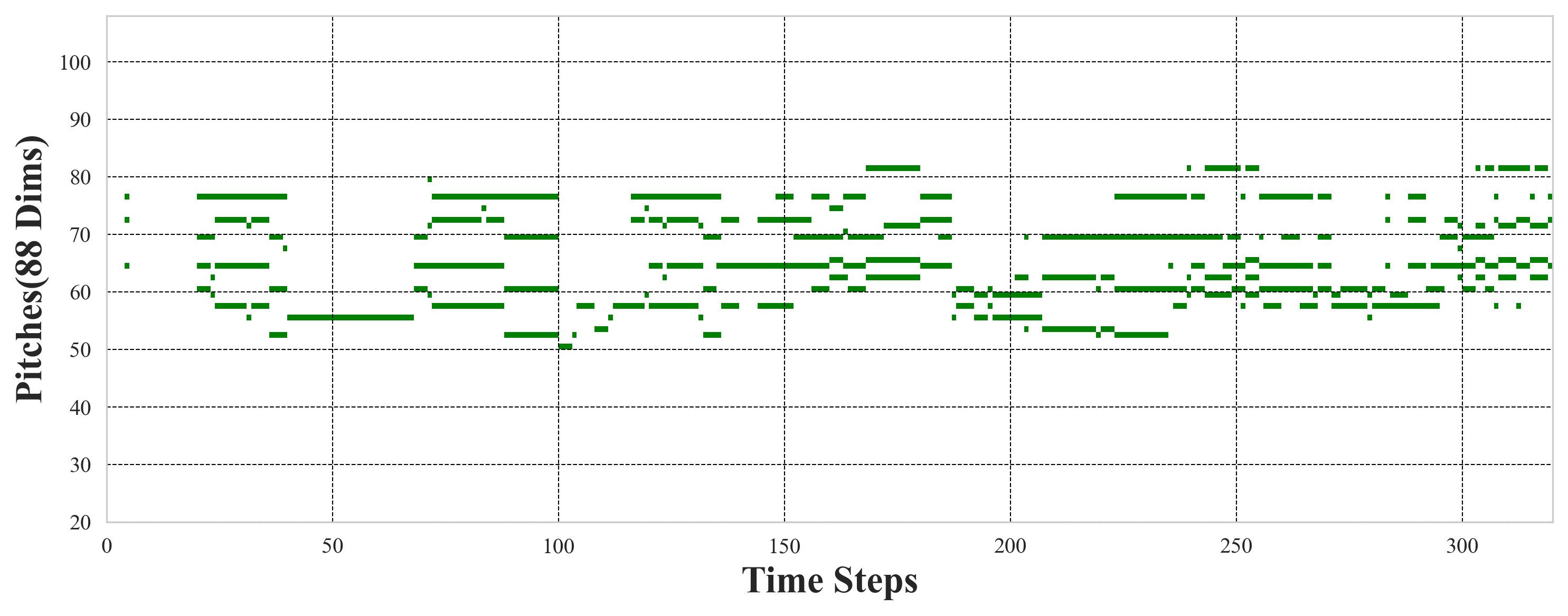}
\caption{Example of 30-second music generated by Hi-ACG.}
\label{fig:gen_30s}
\end{figure*}

\begin{figure*}[t]
\centering
\includegraphics[width=0.9\textwidth]{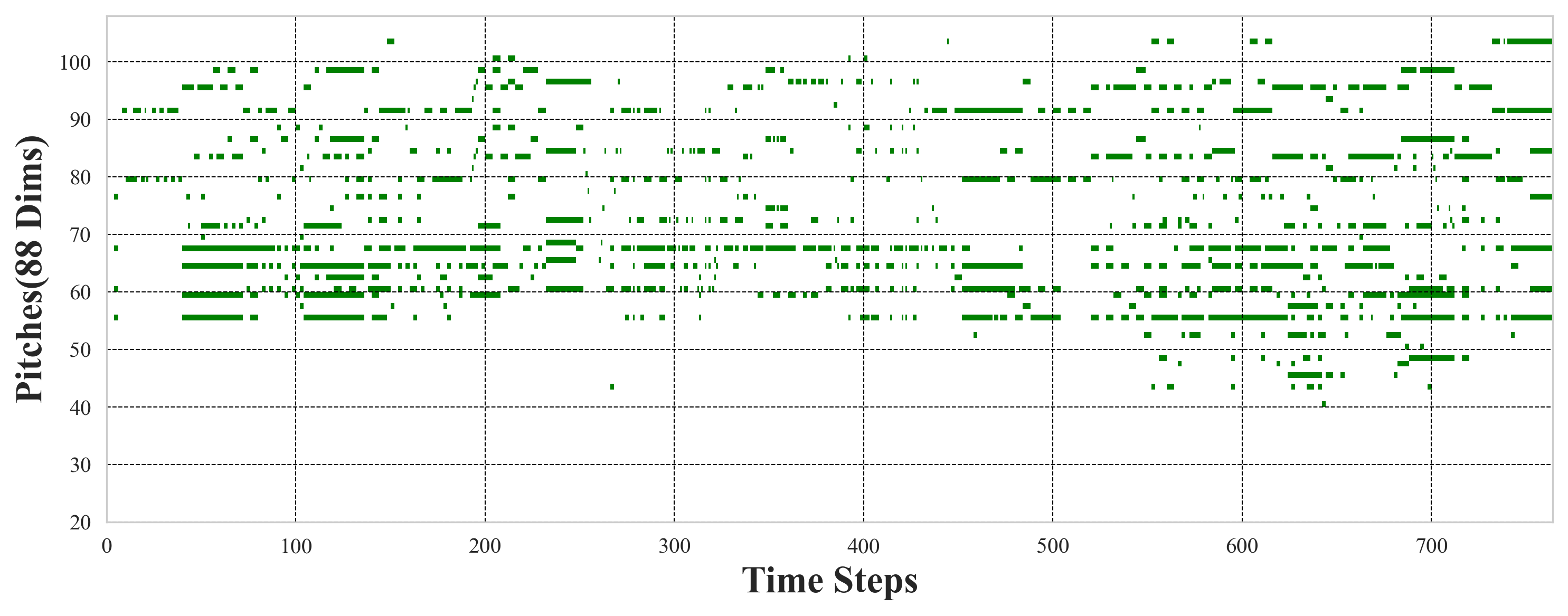}
\caption{Example 1 of 2-minute unconditional music generated by Hi-ACG.}
\label{fig:gen_2m1}
\end{figure*}

\begin{figure*}[t]
\centering
\includegraphics[width=0.9\textwidth]{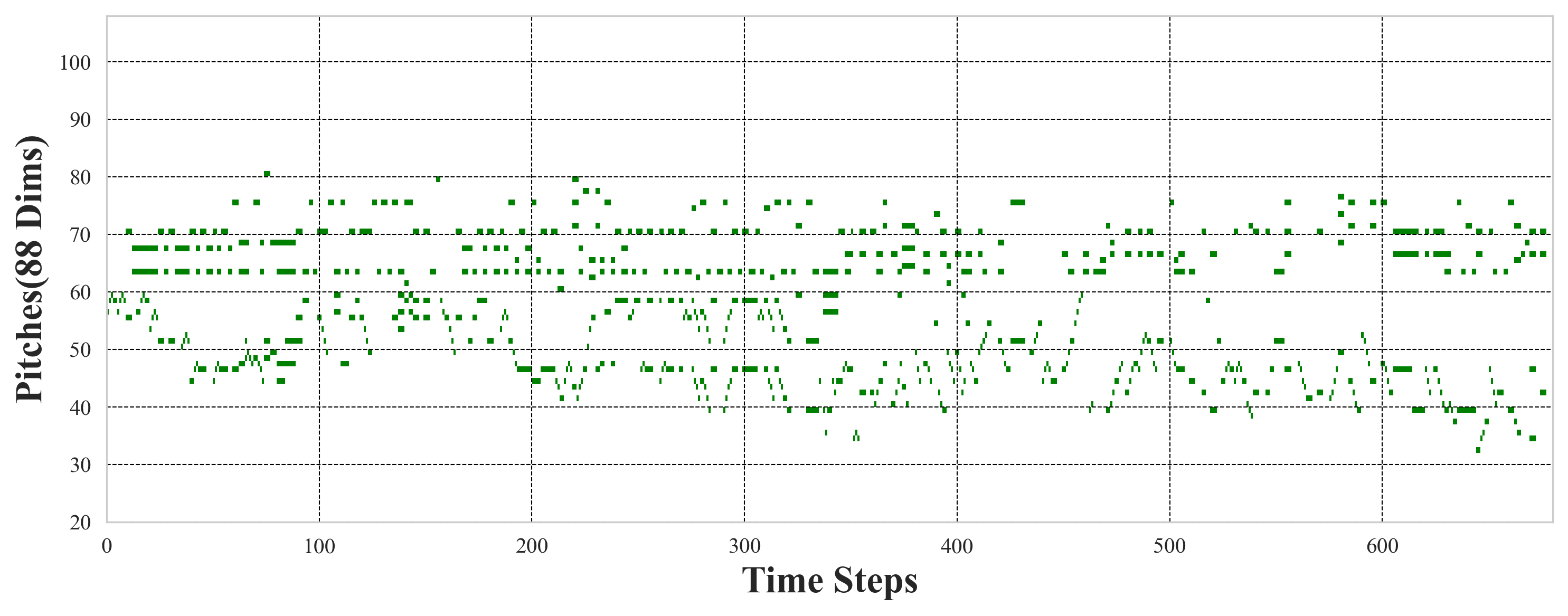}
\caption{Example of 2-minute conditional music generated by Hi-ACG.}
\label{fig:gen_2mc}
\end{figure*}

Figures~\ref{fig:gen_30s} to~\ref{fig:gen_2mc} show piano roll visualizations of music generated by Hi-ACG.

\end{document}